\documentstyle[12pt,fleqn,rotate,epsfig]{article}
\begin{document}
\title{Differential Flow of Protons in $Au+Au$ Collisions at AGS Energies
}
\author{P. K. Sahu$^1$ and W. Cassing$^2$ \\ \\
$^1$Institute of Physics, Bhubaneswar 751005, India\\ $^2$
Institut f\"ur Theoretische Physik, Universit\"at Giessen\\
D-35392 Giessen, Germany }
\maketitle

\abstract{ We study the proton sideward and elliptic differential
flow for $Au+Au$ collisions at AGS energies (2 -- 8 A$\cdot$GeV)
in a microscopic relativistic transport model that includes all
baryon resonances up to a mass of 2 GeV as well as string degrees
of freedom for the higher hadronic excitations. In order to
explore the sensitivity of the various differential flows to the
nuclear equation of state (EoS) we use three different
parameterizations of the scalar- and vector mean-fields, i.e.
 NL2 (soft), NL23 (medium) and NL3 (hard), with their momentum
dependence fitted to the experimental  Schr\"odinger equivalent
potential (at normal nuclear matter density $\rho_0$) up to
kinetic energies of 1 GeV. We calculate the excitation function of
sideward and elliptic flow within these parameter sets for $Au+Au$
collisions and compare with the recent data from the E895
Collaboration as a function of rapidity, impact parameter and
transverse momentum, respectively. We find that the best
description of the differential data is provided by a rather
'stiff' EoS at 2 A$\cdot$GeV (NL3) while at higher bombarding
energies (4--8 A$\cdot$GeV) a 'medium' EoS leads to the lowest
$\chi^2$ with respect to the data. However, the differences in the
transverse and elliptic flows (from the different parameter sets)
become of minor significance at 4--8 A$\cdot$GeV. We attribute
this insensitivity to a similar reduction of the vector potential
in all models and to the dominance of string degrees of freedom at
these bombarding energies. }

\vskip 0.2 in
{PACS: 25.75.+r, 24.10.Jv }

{\noindent
Keywords: Relativistic heavy-ion collisions, relativistic models,
collective flow}

\newpage
\section{Introduction}
%

According to lattice calculations of QCD a phase transition from
an interacting hadronic matter to partonic matter (interacting
quarks and gluons) is expected at temperatures $T_c \approx 150 -
180$ MeV for vanishing quark chemical potential \cite{lattice}.
Recent extrapolations to finite quark chemical potentials $\mu_q
\neq$ 0 suggest a decrease of the critical temperature $T_c$ with
increasing $\mu_q$ \cite{lattice1}. Whereas rather low chemical
potentials ($\sim $ 30 MeV) can be probed in central $Au+Au$
collisions at the full RHIC energy ($\sqrt{s}$ = 200 GeV)
\cite{PBM1,RHIC1}, sizeable chemical potentials or baryon
densities are encountered in nucleus-nucleus collisions at AGS
energies of 2--11 A$\cdot$GeV. Currently one expects the phase
transition at low $\mu_q$ to be of second order
\cite{Shuryak0,Shuryak} and a tricritical point for $\mu_q
\approx$ 230 MeV, $T_c(\mu_q) \approx 160$ MeV \cite{Katz}. Such
conditions might be encountered in $Au+Au$ collisions at AGS
energies leading eventually to a softening of the hadronic
equation of state (EoS) since the pressure in the partonic phase
is expected to be lower than in the hadronic regime. A decrease of
the pressure, furthermore, should go along with a decrease of the
collective flow of hadrons once the phase boundary is met
\cite{Rischke}. Thus a systematic study of flow variables in
energetic nucleus-nucleus collisions might indicate the appearance
of a new phase of hadronic matter \cite{E895a}. We recall, that
the collective flow observable has been exploited for more than a
decade to obtain information on the EoS \cite{cass90} -
\cite{Fuchs}.

In the last years both, the directed transverse flow (sideward
flow) and the flow tensor (elliptic flow), have been measured
\cite{E895a,her96,rei97,E877,E895,E895c,E895d} for heavy-ion
($Au+Au$) collisions at AGS energies in the incident energy range
of 2 A$\cdot$GeV $\le E_{inc} \le 11 $ A$\cdot$GeV. More recently,
also differential flow data have become available from the E985
Collaboration \cite{E895b} which allow for more severe constraints
on the E0S.

The sideward flow is produced at the early stage of
nucleus-nucleus collisions and hence reflects the early pressure
gradients in the collisions. Thus the strength of the momentum per
particle $<p_x>(y)$ in the reaction plane versus rapidity $y$
sheds some light on the initial pressure since the gradients are
dominated by the collision geometry. The elliptic flow emerges
from the early squeeze-out of compressed matter that - depending
on bombarding energy - is shadowed by the spectator nucleons.  At
low energies this squeeze-out of nuclear matter leads to a
negative elliptic flow since projectile and target spectators
distort the collective expansion of the 'fireball' in the reaction
plane. At high energies the projectile and target spectators do
not hinder anymore the in-plane expansion of the 'fireball' due to
their high velocity ($\approx c$); the elliptic flow then becomes
positive. The competition between squeeze-out and in-plane
elliptic flow at AGS energies depends on the nature of the nuclear
force as pointed out by Danielewicz et al. \cite{dani98}.
Furthermore, it has been suggested in Ref. \cite{E895b} that the
elliptic flow is sensitive to the stiffness of the EoS  and
possibly to quark-gluon plasma (QGP) formation.

In Refs. \cite{sahu98,Sahu00,Larionov} both the sideward and
elliptic flow of nucleons has been studied for various
nucleus-nucleus collisions as a function of beam energy from 0.05
A$\cdot$GeV \cite{Larionov} to 11 A$\cdot$GeV \cite{Sahu00}
employing a relativistic transport model with hadronic and string
degrees of freedom. In the AGS energy regime, however, our
calculations have been performed only with a 'stiff' EoS based on
momentum-dependent scalar and vector self energies for the
nucleons. Here we continue our studies within the same transport
approach using, furthermore, different equations of state with
incompressibilities of 210 MeV (soft), 300 MeV (medium), and 380
MeV (hard). Moreover, we will confront our calculations with the
recent differential flow data from the E985 Collaboration
\cite{E895b} as a function of impact parameter, rapidity and
transverse momentum.

We organize our work as follows: In Section 2 we briefly describe
the relativistic transport approach and fix the parameter sets for
the scalar and vector self energies in comparison to the
'experimental' optical potential from Ref. \cite{Ham90}. In
Section 3 we will systematically study the transverse in-plane
flow for $Au+Au$ collisions from 2--8 A$\cdot$GeV while Section 4
concentrates on elliptic flow as a function of impact parameter or
momentum, respectively.  Section 5 concludes this study with a
summary and discussion of open questions.

\section{Description of the transport approach}
Our studies are based on the relativistic hadron-string model used
in Ref.~\cite{Sahu00} for the investigation of transverse and
elliptic baryon flow at AGS energies. The approach  is based on a
coupled set of covariant transport equations for the phase-space
distributions of hadrons which involve the propagation in
momentum-dependent scalar and vector mean fields as well as
transition rates for elastic and inelastic scattering processes.
In the collision terms we employ the in-medium cross sections as
given in Ref. \cite{Effe}, that are parameterized in line with the
corresponding experimental data (in vacuum) for $\sqrt{s} \leq$
3.5 GeV, while for higher invariant energies $\sqrt{s}$ the Lund
string formation and fragmentation model \cite{Lund} is employed
as in  the Hadron-String-Dynamics (HSD) approach
\cite{cass99,ehe96}, which has been used extensively for the
description of particle production in nucleus-nucleus collisions
from SIS to SPS energies \cite{cass99}. We note, that the 'string
threshold' of 3.5 GeV has been determined by a fit to the
transverse mass spectra of protons in central $Au+Au$ collisions
at AGS energies (cf. Ref. \cite{Sahu00}).

The mean-field part of the relativistic transport model is fully
specified by the scalar and vector potentials, which determine the
mean-field propagation of the hadrons. Once these potentials are
specified, the energy per particle $E/A$ as a function of density
$\rho$ (EoS) can be calculated for nuclear matter ground-state
configurations in a straight forward manner \cite{mar94,lan91}. We
recall that in a simple relativistic mean-field theory one assumes
the scalar and vector fields to be represented by point-like
meson-baryon (i.e. momentum independent) couplings. Such models
explain the Schr{\"o}dinger-equivalent potential (nucleon optical
potential) at low kinetic energies ($E_{kin} \leq 200$ MeV) in
nuclear matter also in accordance with data \cite{Ham90} and more
sophisticated Dirac-Brueckner calculations \cite{Mal,Fu1,Fu2}.
However, at higher kinetic energies (200 MeV $\leq E_{kin} \leq$ 1
GeV), this local relativistic mean-field theory does no longer
describe the nucleon optical potential - as extracted from elastic
$p+A$ scattering \cite{Ham90} - due to its linear function in
$E_{kin}$. Since the energy dependence of sideward flow and
elliptic flow are controlled in part by the nucleon optical
potential, a local relativistic mean-field model cannot be applied
anymore to high-energy heavy-ion collisions \cite{sahu98}. In
order to take care of this aspect we invoke an explicit
momentum-dependence of the coupling constants, i.e. a form factor
for the meson-baryon couplings \cite{cass99,ehe96} as also used in
\cite{Sahu00}.

In the present calculations we consider  Lagrangian densities with
nucleon, $\sigma$ (scalar) and $\omega$ (vector)  fields and
nonlinear self-interactions of the scalar field as in
\cite{Sahu00}, however, with different parameter sets. These
parameter sets are denoted by NL2 (incompressibility K=210 MeV,
soft), NL23 (K=300 MeV, medium) and NL3 (K=380 MeV, hard)
according to the stiffness of the EoS,
respectively\cite{SahuNS,lan91}. For all these parameter sets
scalar and vector form factors at the vertices are introduced in
the form \cite{Sahu00}
\begin{equation}
\label{form}
    f_s({\bf p})=\frac{\Lambda_s^2-a{\bf p}^2}{\Lambda_s^2+{\bf p}^2}
  \qquad\mbox{and}\qquad
    f_v({\bf p})=\frac{\Lambda_v^2-b{\bf p}^2}{\Lambda_v^2+{\bf p}^2}.
\end{equation}
In (\ref{form}) the cut-off parameters $\Lambda_s$ and $\Lambda_v$
and constants  $a$ and $b$ are obtained by fitting the
Schr\"odinger equivalent potential,
\begin{equation}
\label{pot}
U_{sep} (E_{kin}) = U_s + U_0 + \frac{1}{2M} (U_s^2-U_0^2) +
\frac{U_0}{M} E_{kin},
\end{equation}
to Dirac phenomenology for intermediate energy proton-nucleus
scattering \cite{Ham90}.
These parameters $\Lambda_s$, $\Lambda_v$, $a$ and $b$ are 1.05
GeV, 1.52 GeV, 1/3 and 1/4 for NL2, 1.0 GeV, 1.35 GeV, 1/3 and 1/4
for NL23 and 1.0 GeV, 0.9 GeV, 1/2 and 1/6 for NL3, respectively.
The above momentum dependence is computed self-consistently on the
mean-field level as  in Ref. \cite{Sahu00}.

Fig. 1 displays the resulting Schr\"odinger equivalent potential
(\ref{pot}) at density $\rho_0$ = 0.168 fm$^{-3}$ as a function of
the nucleon kinetic energy $E_{kin}$ with respect to the nuclear
matter at rest for all three models NL2 (solid line), NL23 (dotted
line) and NL3 (dot-dashed line). These models are compared with
the data from Hama et al. \cite{Ham90} (full squares) which are
well described due to the explicit fit. Since the data are only
available up to 1 GeV our extrapolations via (\ref{form}) to
higher kinetic energies have to be taken with care. However, as
shown in Refs. \cite{sahu98,Sahu00} a decrease of $U_{sep}$ with
energy is necessary to properly describe the transverse flow of
nucleons in heavy-ion reactions at AGS energies. We mention, that
the form factors (\ref{form}) lead to negative optical potentials
for high energies which we consider as unphysical; for simplicity
we have set the potentials to zero above 3.2 GeV, 2.8 GeV and 3.5
GeV for NL2 (solid line), NL23 (dashed line) and NL3 (dot-dashed
line), respectively, as illustrated in Fig. 1.

It is interesting to note that in order to match the Schr\"odinger
equivalent potentials up to 1 GeV kinetic energy, the strength of
the vector potential (with a momentum-dependent form factor) for
the soft equation of state (NL2) is higher than for the stiffer
equations of state (NL23 and NL3). Since the vector potentials
must be partly compensated by the attractive scalar potentials
according to (\ref{pot}) this also holds true for the strength of
the scalar potentials.

The energy per nucleon $E/A$ - as resulting from the 3 parameter
sets - is shown in the upper left part of Fig. 2 as a function of
the density $\rho$, where the free nucleon  mass has been
subtracted. The shorthand notations 'soft', 'medium' and 'stiff'
become obvious from this figure. It is of further importance, how
the optical potentials look like as a function of the baryon
momentum $p$ with respect to the nuclear matter at rest e.g. for
densities of  2 $\rho_0$, 3 $\rho_0$, and 5 $\rho_0$ which are
encountered in nucleus-nucleus collisions at AGS energies. This
information is also displayed in Fig. 2 for NL2 (solid lines),
NL23 (dotted lines) and NL3 (dot-dashed lines). Whereas for low
momentum $p$ the optical potential $U_{sep}$ becomes more
repulsive with increasing stiffness $K$ (and density $\rho$), it
is of roughly the same size at all densities for $p \approx$ 1
GeV/c and even most repulsive for the 'soft' parameter set NL2
above $\sim$1.5 GeV/c. Thus the energy per particle $E/A$ in the
upper left part of Fig. 2 is somewhat misleading since it only
reflects the momentum dependence of the potentials up to the Fermi
momentum
\begin{equation}
\label{pf}
 p_F = (\frac{3}{2} \pi^2 \rho)^{1/3},
\end{equation}
which is less than 0.5 GeV/c even at 6 $\rho_0$.

Some further note of caution has to be added here concerning the
interpretation of 'baryons per volume', i.e. the density $\rho$,
since in the transport model the initial high density phase is
described not by 'formed' baryons but by 'strings', that
correspond to continuum excitations of the hadrons. Thus, when
addressing the baryon density, we actually consider the
constituent quark number per volume - as contained in the strings
- and divide by 3 in order to define the baryon density $\rho$. As
shown explicitly in Fig. 11 of Ref. \cite{Cass02} for central
$Au+Au$ reactions at 11.6 GeV/c, the initial high density phase of
the reaction is fully dominated by quarks and diquarks that form
the endpoints of the strings. In this respect we compute the quark
current $j_\mu(t,{\bf r})$ and define the 'baryon' density as
\begin{equation}
\rho(t,{\bf r})= \rho(x) = \frac{1}{3} \sqrt{j_\mu(x) j^{\mu}(x)}.
\end{equation}

\section{Transverse flow}


In this Section we restrict ourselves to the transverse in-plane
flow $<p_x>(y)$ for mid-peripheral $Au+Au$ collisions from 2--8
A$\cdot$GeV. Since the 'input' of the transport approach has been
presented in the previous Section we can directly step on with the
numerical results in comparison to the experimental data
available.

In Fig. 3 we display the average proton in-plane momentum
$<p_x>(y^\prime)$ as a function of the normalized rapidity
\begin{equation}
y^{\prime} = \frac{y_{cm}}{y_{proj}} \end{equation}
for $Au+Au$
systems at 2--8 A$\cdot$GeV  for the impact parameter $b=6$ fm.
Since experimentally the sideward flow is observed for
mid-peripheral collisions with an average impact parameter of
$b=6$ fm, we restrict to a single impact parameter here. Note,
that the flow only marginally changes with $b$ for mid-peripheral
reactions.
The solid lines in Fig. 3 correspond to the parameter set NL2
(soft), the dashed lines represent the set NL23 (medium) and the
dot-dashed lines the set NL3 (hard). The full squares are the
experimental data from the E895 Collaboration \cite{E895a}. The
direct measurements have been performed for 0 $\leq y^\prime \leq$
1 and the data for -1 $\leq y^\prime \leq$ 0 have been generated
by reflection around $y^\prime$ = 0. We observe that the overall
description of the data is reasonably good for all parameter sets.
To quantify the 'agreement' we have performed a $\chi^2$ analysis
(for -0.6 $\leq y^\prime \leq $ 0.6) and find that the best
description is given by the NL3 model at 2 A$\cdot$GeV, while from
4-8 A$\cdot$GeV the 'medium' NL23 parameter set performs best. One
might interpret this as an indication for a 'softening' of the EoS
at $\sim$ 4 A$\cdot$GeV, however, the differences in the $\chi^2$
values are only small such that this trend has to be interpreted
with care.

The shape of $<p_x(y^{\prime})>$ is, furthermore, fitted by a
polynomial function \begin{equation} \label{fit} <p_x(y^{\prime})>
\approx F y^{\prime}+Ky^{\prime 3} \end{equation} over a finite
interval centered around midrapidity. The linear coefficient $F$
in (\ref{fit}) is denoted as sideward flow $F$. This quantity is
shown in Fig. 4 for the parameter sets NL2 (solid line),  NL23
(dotted line) and NL3 (dot-dashed line) in comparison to the data
from Ref. \cite{E895a} (full squares). We find that all parameter
sets are roughly compatible with the data such that the transverse
flow observable $F$ does not qualify very much for a determination
of the stiffness of the EoS as pointed out in Ref. \cite{cass90}
more than a decade ago. Only at 2 A$\cdot$GeV the data more
clearly favor a 'stiff' EoS as provided by NL3. The reasonable
description of the transverse flow $F$ at 4-8 A$\cdot$GeV by all
parameter sets is attributed to the fact that the momentum and
density  dependence of the scalar and vector mean fields is
roughly the same for momenta $p \geq$ 0.5 GeV/c (cf. Figs. 1 and
2), which is enforced by the fit to the experimental data from
Hama et al. \cite{Ham90}.

The significant decrease in $F$ for bombarding energies above 2
A$\cdot$GeV is due to the reduction of the vector potentials with
momentum as pointed out before in Ref. \cite{Sahu00}. According to
our understanding the reduction of the vector coupling with
density and momentum is a genuine phenomenon of nuclear many-body
physics and points towards a restoration of chiral symmetry with
increasing density as advocated by Brown and Rho \cite{Gerry}; it
not necessarily has to be interpreted as a signature for a phase
transition to a QGP.

\section{Elliptic flow}
We now turn to the results of our calculations for the elliptic
flow $v_2(b,p_t)$. The elliptic flow can be measured by  the
second Fourier coefficient of the azimuthal distribution of
particles with respect to the reaction plane and is characterized
by the expectation value \cite{Star01},
\begin{equation}
\label{v2} v_2=<({p_x}^2-{p_y}^2)/(({p_x}^2+{p_y}^2)>.
\end{equation}
In Fig. 5 we show the impact parameter $b$ dependence of the
elliptic flow (\ref{v2}) for $Au+Au$ collisions at energies of 2,
4 and 6 A$\cdot$GeV without any cut on the transverse momentum
$p_t$ of the protons. As before, the solid lines display the
results from the set NL2, the dotted lines those from NL23 and the
dot-dashed lines those from NL3. The experimental data (full
squares) have been taken from Ref. \cite{E895b}. We note that the
impact parameter determination from experiment only holds within
$\pm$ 1 fm. At 2 A$\cdot$GeV we observe a clear sensitivity to the
stiffness of the EoS; only the set NL3 (stiff) describes the large
negative elliptic flow as a function of $b$ correctly, whereas the
other parameter sets turn out too low in $v_2$. This finding
agrees with that from Danielewicz \cite{dani98} that the 2
A$\cdot$GeV data for $Au+Au$ 'need' a stiff EoS. However, contrary
to Ref. \cite{dani98} we do not find convincing indications for a
softening of the EoS at higher bombarding energy. The $\chi^2$
fits to the data in Fig. 5 give only a tiny preference to the
results from NL23 (medium) at 6 A$\cdot$GeV, whereas NL3 (stiff)
and NL23 (medium) practically have the same $\chi^2$ at 4
A$\cdot$GeV.

The situation might change when looking at the elliptic flow at
high transverse momentum $p_t$. We thus compare in Fig. 6 our
calculations on the elliptic flow as a function of impact
parameter $b$ with the respective data from Ref. \cite{E895b} for
a cut in $p_t \geq$ 0.7 GeV/c for $Au+Au$ at 2 and 4 A$\cdot$GeV.
Again clearly the best description at 2 A$\cdot$GeV is given by
NL3 (stiff) now even slightly underestimating the magnitude of
$v_2$. On the other hand, our results for the soft EoS (given by
NL2) give the wrong sign of $v_2$ at 4 A$\cdot$GeV which in the
experiment is small, but negative at high $p_t$. At 4 A$\cdot$GeV
the best $\chi^2$ is provided by NL23 (medium), but the sign of
$v_2$ comes out wrong for 6 fm $\leq b \leq $ 8 fm, where the
elliptic flow is described best by NL3 (stiff). We thus do not
find a convincing indication for a softening of the EoS at
4A$\cdot$GeV.

We continue with a comparison of our calculations to the data from
\cite{E895b} for the elliptic flow at 6 A$\cdot$GeV imposing a
high $p_t$ cut of 1 GeV/c in Fig. 7. Within experimental error
bars now all parameter sets show results that are compatible with
the data. A $\chi^2$ analysis gives a slight preference for NL23
(medium) followed by NL2 (soft), whereas the $\chi^2$ is higher
for NL3 (stiff). However, these differences are still within the
statistical accuracy of the transport calculations which becomes
poor for high transverse momentum protons. We attribute this
approximate insensitivity to the stiffness of the EoS to the fact,
that the optical potentials at high density and momenta $p \geq$
0.5 GeV are roughly the same (cf. Fig. 2) and that multi-meson
production channels from string decays start to dominate the
reaction dynamics above 4 A$\cdot$GeV as pointed out before in
Ref. \cite{Sahu00}.

\section{Summary and Discussion}
In summary, we have analyzed the sideward and elliptic flow from
$Au+Au$ collisions at beam energies from 2 A$\cdot$GeV to 8
A$\cdot$GeV at AGS energies using different nuclear forces that
involve 'soft', 'medium' and 'stiff' equation of states in the
microscopic relativistic transport model.

The measured sideward flow or the average in-plane momentum as a
function of rapidity can be described in the dynamical transport
model by practically all parameter sets thus demonstrating that
the transverse flow is rather insensitive to the stiffness of the
EoS, but crucially depends on the momentum dependence of the mean
fields  \cite{cass90}. The rather good description of the
transverse flow $F$ by all parameter sets in turn is attributed to
the fact that the momentum dependence of the scalar and vector
mean fields is roughly the same (cf. Figs. 1 and 2), which is
enforced by the fit to the experimental data from Hama et al.
\cite{Ham90}. A $\chi^2$ analysis here favors a 'stiff' EoS at 2
A$\cdot$GeV while the 4-8 A$\cdot$GeV data are best reproduced by
the 'medium' parameter set NL23. However, the significance for
this trend is very low.

Our calculations for the elliptic flow $v_2$ (\ref{v2}) show a
clear sensitivity to the stiffness of the EoS at 2 A$\cdot$GeV;
only the set NL3 (stiff) describes the large negative elliptic
flow seen by E895 \cite{E895b} as a function of $b$ correctly,
whereas the other parameter sets turn out too low. This finding
agrees with that from Danielewicz \cite{dani98} that the 2
A$\cdot$GeV data for $Au+Au$ call for a stiff EoS. However,
contrary to Ref. \cite{dani98} we do not find convincing
indications for a softening of the EoS at higher bombarding energy
even when gating on high transverse momenta ($p_t \geq$ 0.7 GeV/c
or 1 GeV/c, respectively). A $\chi^2$ analysis favors again the
'medium' EoS (given by the set NL23) at 4-6 A$\cdot$GeV in line
with the analysis of the transverse flow, but with rather low
significance. For bombarding energies $\geq$ 6 A$\cdot$GeV our
calculations turn out to be almost insensitive to the stiffness of
the EoS, which is likely due to the fact that the optical
potentials at high density and momenta $p \geq$ 0.5 GeV (in the
transport approach) are roughly the same (cf. Fig. 2) and that
multi-meson production channels from string decays start to
dominate the reaction dynamics above 4 A$\cdot$GeV \cite{Sahu00}.

The open questions -- and uncertainties in the transport approach
-- are related to the explicit momentum dependence of the optical
potential $U_{sep}$ (\ref{pot}) in the kinematical range above 1
GeV of relative kinetic energy. Here data from elastic
proton-nucleus scattering at AGS would be of significant help to
pin down the present uncertainties. Furthermore, the question of
the 'effective degrees of freedom' encountered in the early phase
of nucleus-nucleus collisions at AGS energies is open, too. The
fact, that our transport calculations -- involving all hadrons up
to a mass of 2 GeV and strings for the hadronic continuum -- give
a  good description of the differential sideward and elliptic flow
data, does not imply that the proper degrees of freedom are
employed. On the other hand, we can turn the argument around and
conclude, that the differential flow data do not necessarily
indicate the presence of a new state of matter such as the QGP.
Electromagnetic signals (hard photons, dileptons or 'penetrating'
probes) should provide complementary information to the collective
observables studied here. Unfortunately, such information is not
available from the AGS; related studies are foreseen at the future
GSI upgrade \cite{GSI}.

%


\newpage

\begin{figure}
\centerline{~{\psfig{figure=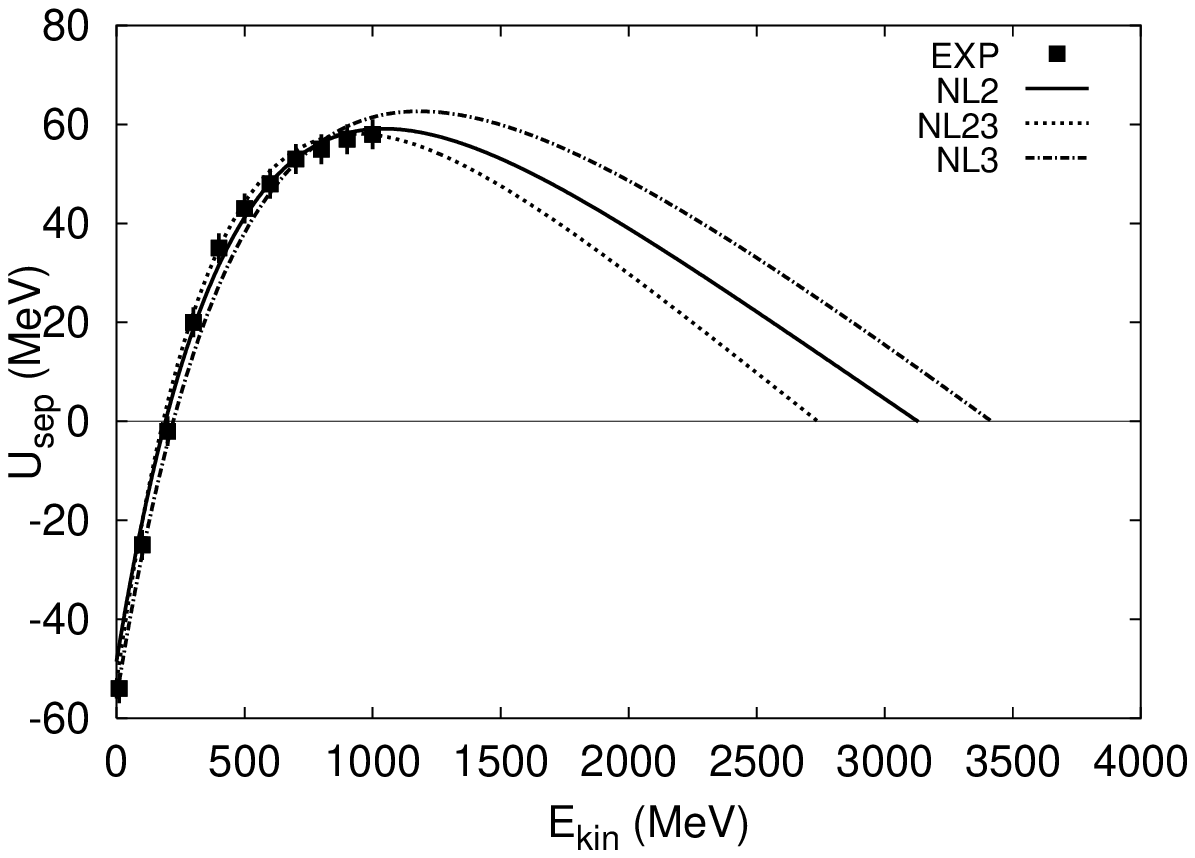,width=14cm,angle=0}}~}
\caption{The Schr\"odinger equivalent potential (2) at density
$\rho_0$ as a function of kinetic energy in comparison to the data
are from~\cite{Ham90} (full squares). The solid line results from
the parameter set NL2 (soft) while the dotted and dot-dashed lines
correspond to NL23 (medium) and NL3 (stiff), respectively. }
\label{Fig:sep}
\end{figure}

\begin{figure}
\centerline{~{\psfig{figure=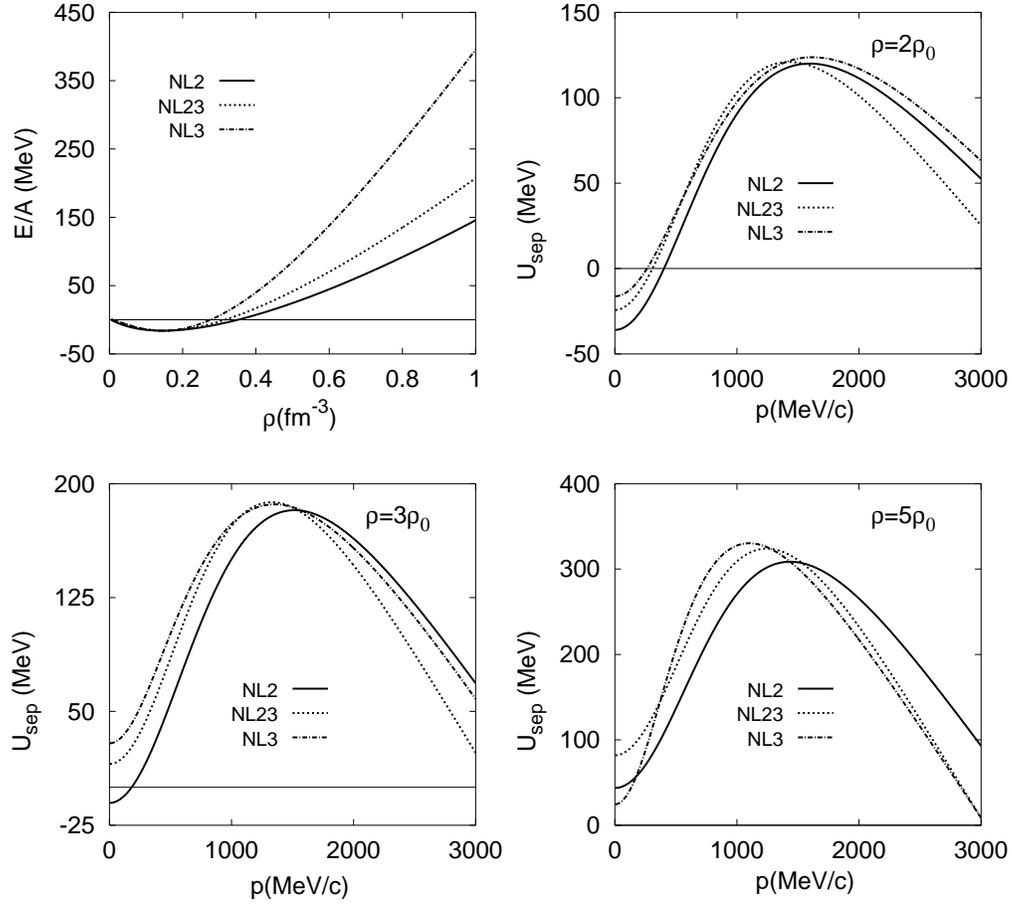,width=14cm,angle=0}}~}
\caption{The energy per nucleon $E/A$ as a function of density
$\rho$ is shown in the upper left part for the parameter sets NL2
solid line), NL23 (dotted line) and NL3 (dot-dashed line).  The
optical potential (2) as a function of the baryon momentum $p$
with respect to the nuclear matter at rest frame  is displayed in
the upper right  and lower parts for densities of 2 $\rho_0$, 3
$\rho_0$ and 5 $\rho_0$, respectively.} \label{Fig:fig2}
\end{figure}

\begin{figure}
\centerline{~{\psfig{figure=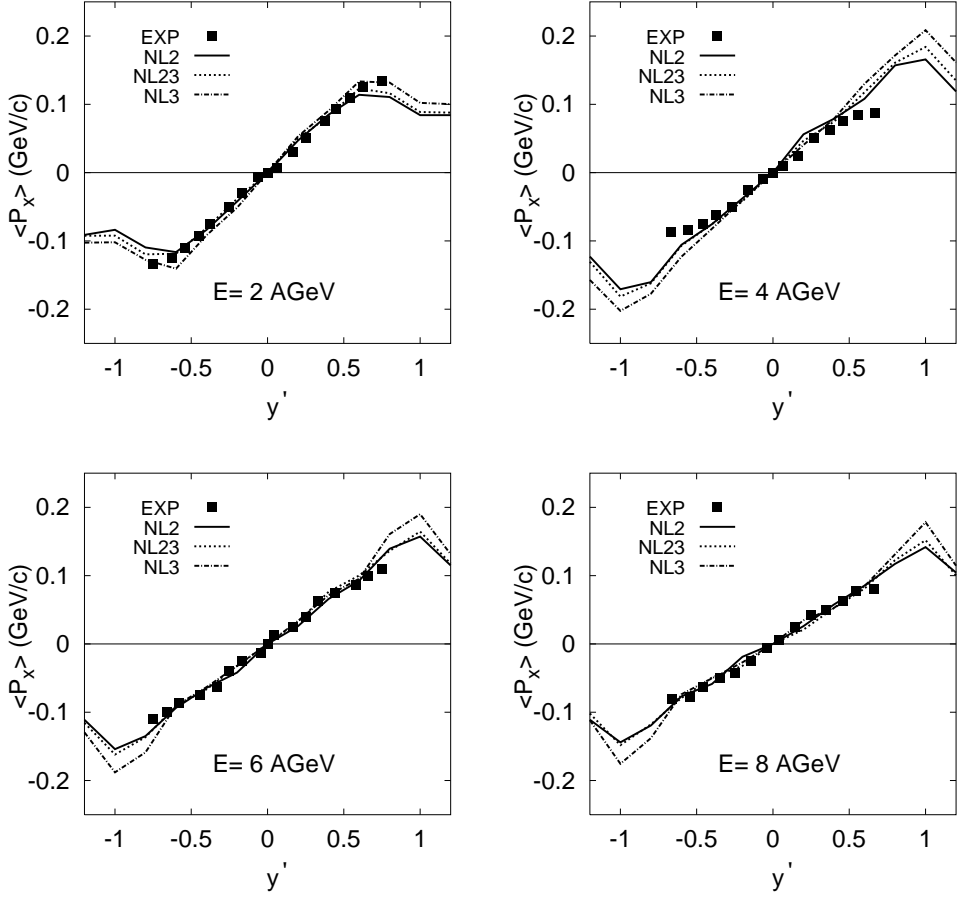,width=14cm,angle=0}}~}
\caption{The average proton in-plane momentum as a function of the
normalized rapidity $y^\prime$ (5) for $Au+Au$  at 2--8
A$\cdot$GeV and impact parameter $b=6$ fm for the parameter sets
NL2 (solid lines), NL23 (dotted lines) and NL3 (dot-dashed lines).
The experimental data (full squares) are from the E895
Collaboration \protect\cite{E895a}.  } \label{Fig:fig3}
\end{figure}

\begin{figure}
\centerline{~{\psfig{figure=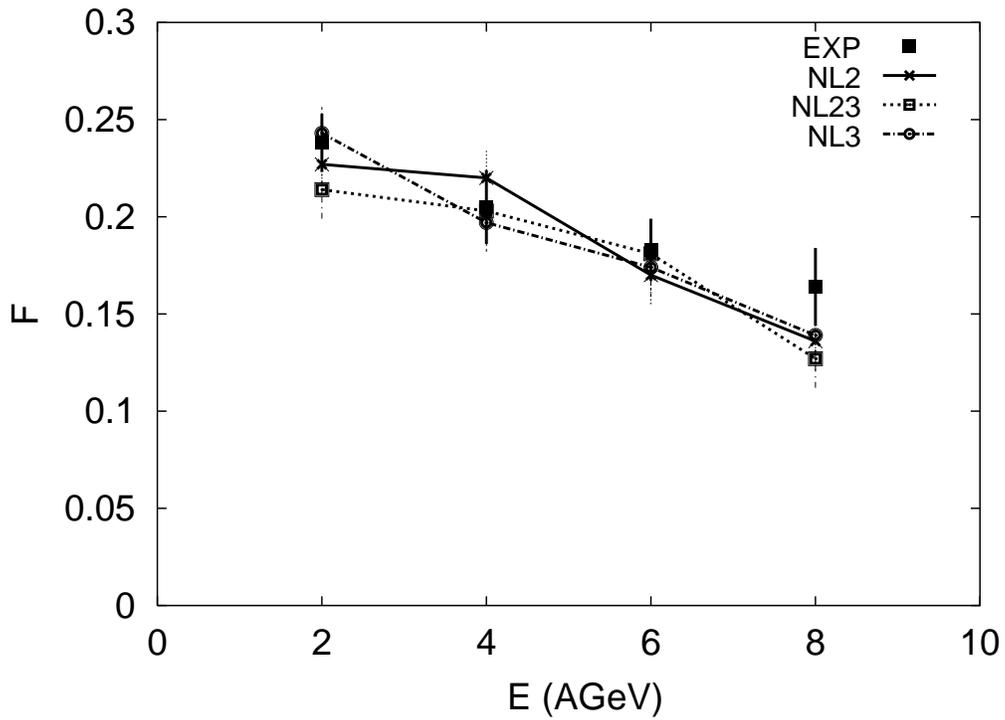,width=14cm,angle=0}}~}
\caption{The flow $F$ (6) as a function of beam energy for $Au+Au$
collisions at b=6 fm for the parameter sets NL2 (solid line), NL23
(dotted line) and NL3 (dot-dashed line). The experimental data
(full squares) are from the E895 collaboration \cite{E895a}.}
\label{Fig:fig4}
\end{figure}

\begin{figure}
\centerline{~{\psfig{figure=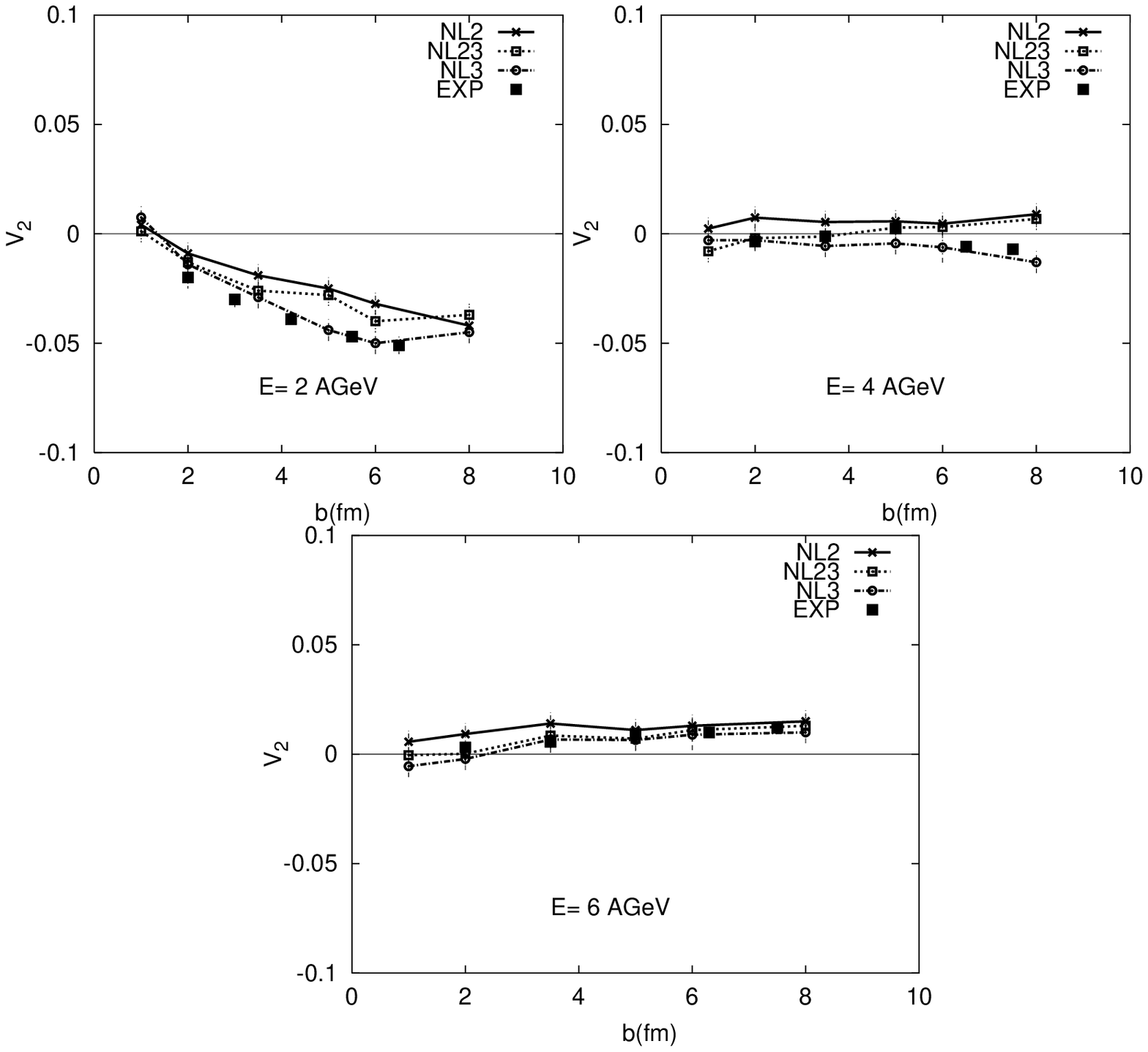,width=14cm,angle=0}}~}
\caption{Elliptic flow $v_2$ (7) as a function of impact parameter
$b$  for $Au+Au$ collisions at 2, 4 and 6 A$\cdot$GeV  with $p_t
\geq 0$. The experimental data (full squares) are from the E895
Collaboration \cite{E895b}.} \label{Fig:fig5}
\end{figure}

\begin{figure}
\centerline{~{\psfig{figure=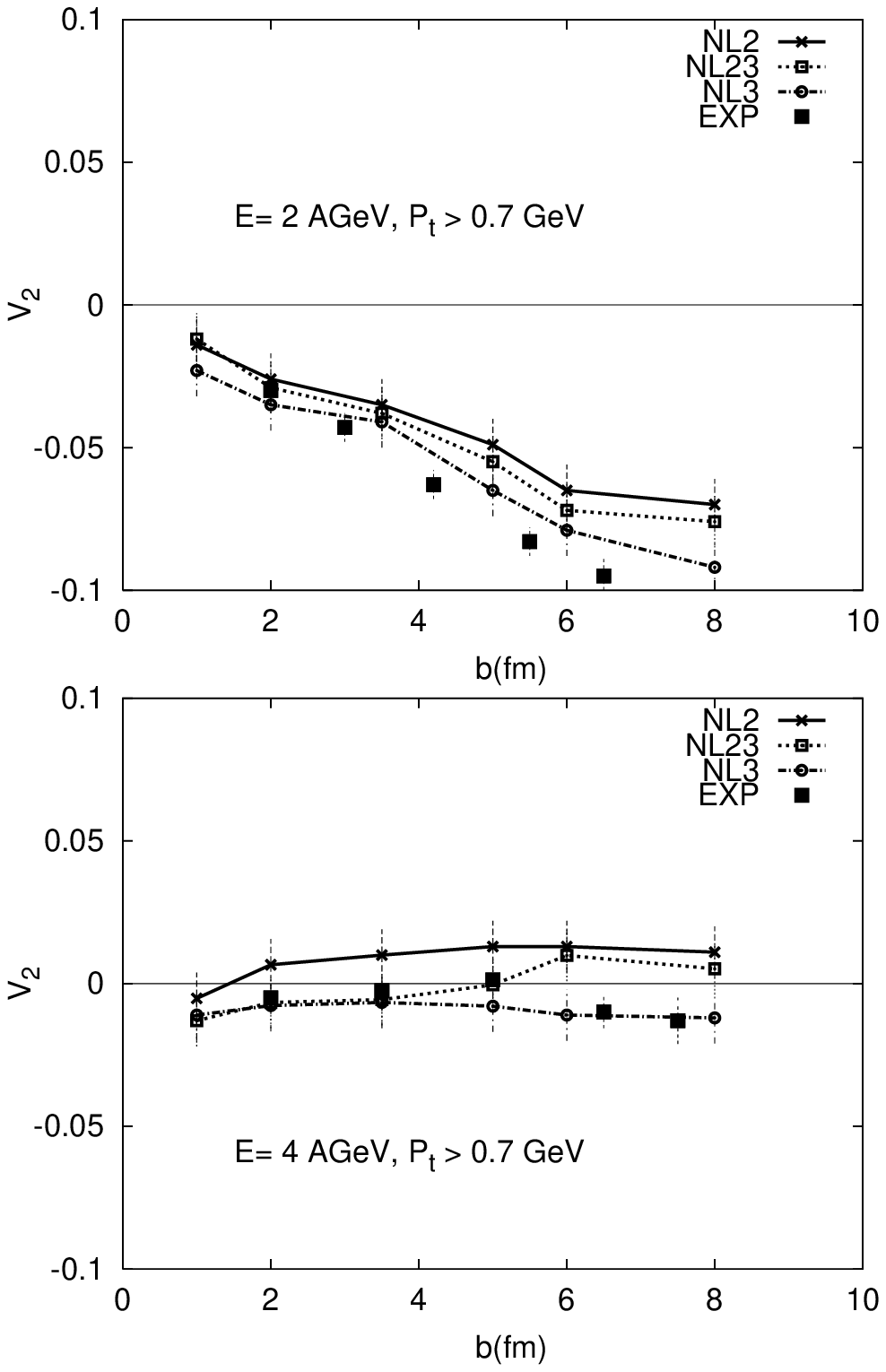,width=14cm,angle=0}}~}
\caption{Elliptic flow $v_2$ (7) as a function of impact parameter
$b$  for $Au+Au$ collisions at 2  and 4 A$\cdot$GeV  for $p_t \geq
0.7$ GeV/c. The experimental data (full squares) are from the E895
Collaboration \cite{E895b}. } \label{Fig:fig6}
\end{figure}

\begin{figure}
\centerline{~{\psfig{figure=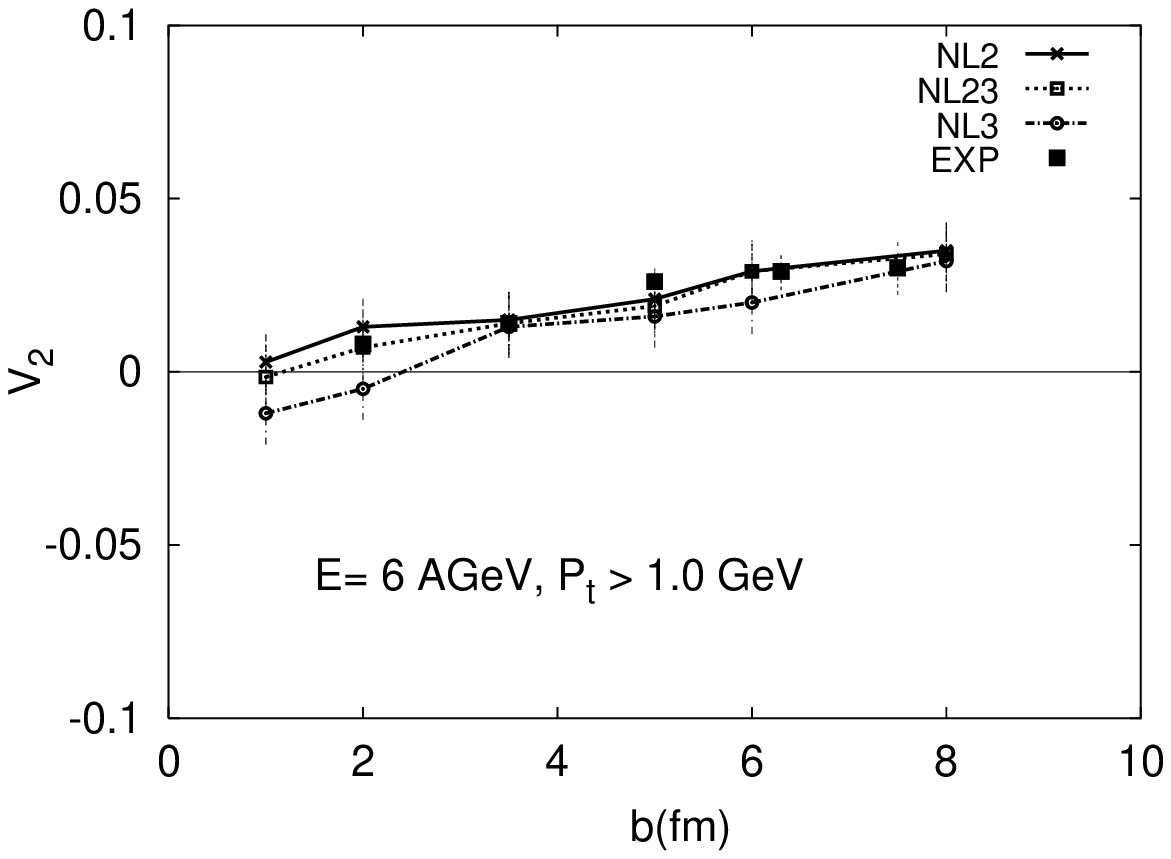,width=14cm,angle=0}}~}
\caption{Elliptic flow $v_2$ (7) as a function of impact parameter
$b$  for $Au+Au$ collisions at 6 A$\cdot$GeV  for $p_t \geq 1.0$
GeV/c. The experimental data (full squares) are from the E895
Collaboration \cite{E895b}.} \label{Fig:fig7}
\end{figure}

\end{document}